\documentclass[final, runningheads]{llncs}
\usepackage[utf8]{inputenc} 
\usepackage{graphicx}
\usepackage[usenames,dvipsnames]{xcolor}
\usepackage{mathtools}
\usepackage{subcaption}
\graphicspath{{./figs/}}

\usepackage[hidelinks]{hyperref}
\usepackage{lineno}

\spnewtheorem{conjct}{Conjecture}{\bfseries}{\itshape}
\usepackage{thmtools}

\renewcommand{\orcidID}[1]{\href{https://orcid.org/#1}{\includegraphics[scale=.03]{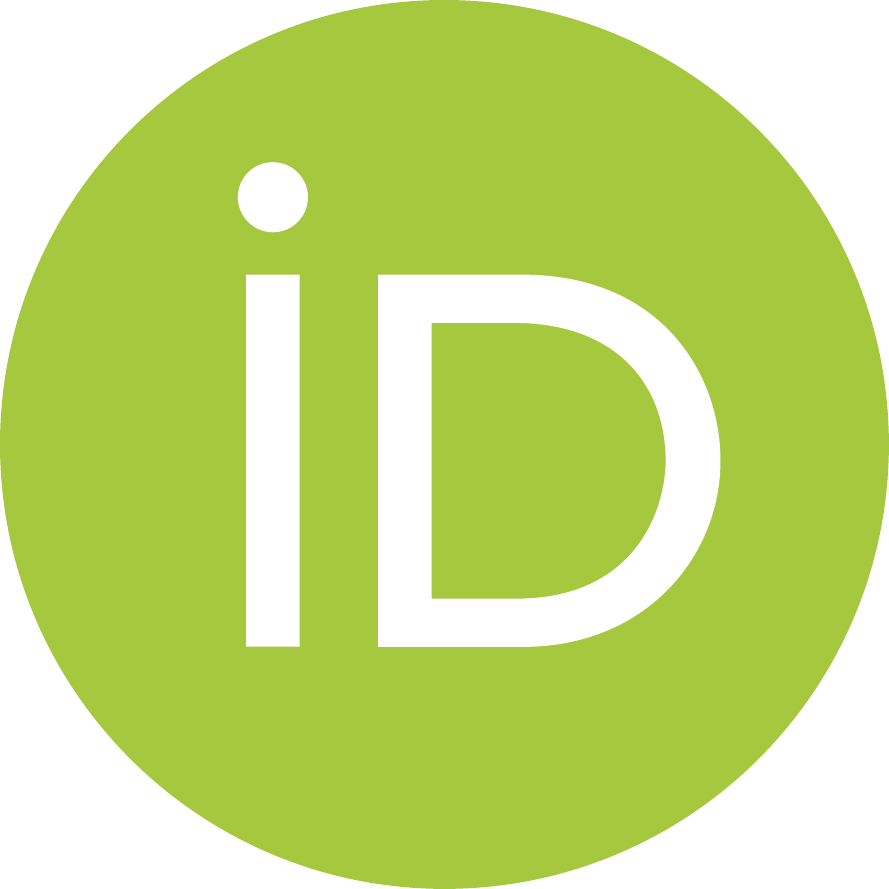}}}

\usepackage{xspace}
\newcommand{\cmonotone}{c-monotone\xspace}

\newcommand{\gtwisted}{generalized twisted\xspace}

\newcommand\blfootnote[1]{%
  \begingroup
  \renewcommand\thefootnote{}\footnote{#1}%
  \addtocounter{footnote}{-1}%
  \endgroup
}

\title{Empty Triangles in Generalized Twisted Drawings of~$K_n$
}
	
	\author{%
		Alfredo~Garc\'ia\inst{1}\orcidID{0000-0002-6519-1472}
		\and
		Javier~Tejel\inst{1}\orcidID{0000-0002-9543-7170}
		\and
		Birgit~Vogtenhuber\inst{2}\orcidID{0000-0002-7166-4467}
		\and\\
		Alexandra~Weinberger\inst{2}\orcidID{0000-0001-8553-6661}
}

\authorrunning{Garc\'ia, Tejel, Vogtenhuber, and Weinberger}
\institute{%
Departamento de M\'etodos Estad\'isticos and IUMA, Universidad de Zaragoza, Zaragoza, Spain\\
\email{\{olaverri,jtejel\}@unizar.es} \and
Institute of Software Technology, Graz University of Technology, Graz, Austria\\
\email{\{bvogt,weinberger\}@ist.tugraz.at}
}
\begin{document}

\maketitle

\begin{abstract}
Simple drawings are drawings of graphs in the plane or on the sphere such that vertices are distinct points, edges are Jordan arcs connecting their endpoints, and edges intersect at most once (either in a proper crossing or in a shared endpoint). Simple drawings are generalized twisted if there is a point~$O$ such that every ray emanating from~$O$ crosses every edge of the drawing at most once and there is a ray emanating from~$O$ which crosses every edge exactly once.
We show that all generalized twisted drawings of $K_n$ contain exactly $2n-4$ empty triangles, by this making a substantial step towards proving the conjecture that this is the case for every simple drawing of $K_n$.
	\blfootnote{\hspace{-10pt}\begin{minipage}[l]{0.2\textwidth} \vspace{-8pt}\includegraphics[trim=10cm 6cm 10cm 5cm,clip,scale=0.15]{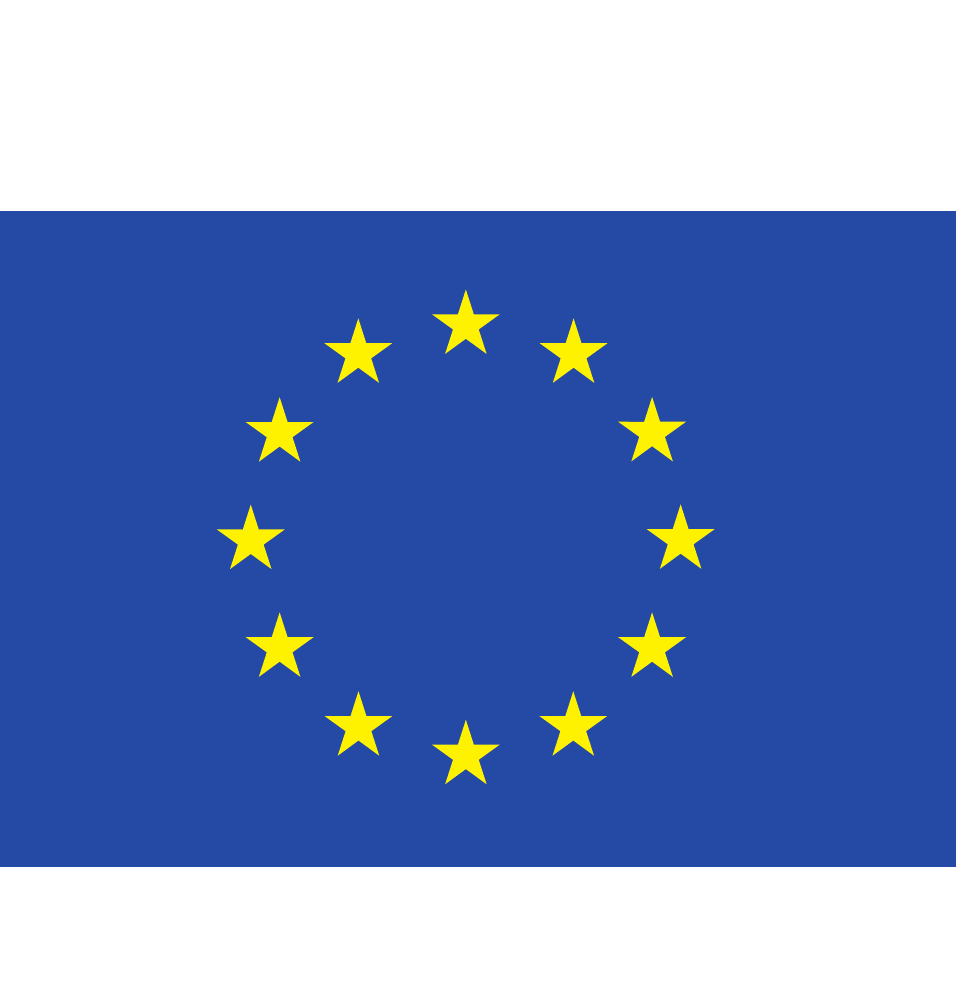} \end{minipage}  \hspace{-1.1cm} \begin{minipage}[l][1cm]{0.87\textwidth} \vspace{-2pt}
      This project has received funding from the European Union's Horizon 2020 research and innovation programme under the Marie Sk\l{}odowska-Curie grant agreement No 734922.
\end{minipage}}
	 \blfootnote{\hspace{-10pt}\begin{minipage}[l]{\textwidth}\footnotesize{
		A.G.\ and J.T.\ partially supported by the Gobierno de Aragón project \mbox{E41-17R.}
		J.T.\ partially supported by project PID2019-104129GB-I00 / AEI / 10.13039/501100011033 of the Spanish Ministry of Science and Innovation.
		B.V.\ partially supported by the Austrian Science Fund as FWF project~\mbox{I 3340-N35} within the collaborative DACH project \emph{Arrangements and Drawings}
		A.W.\ partially supported by the FWF project~W1230.}
	 \end{minipage}}
\end{abstract}

\section{Introduction}
\emph{Simple drawings} are drawings of graphs in the plane or on the sphere such that vertices are distinct points, edges are Jordan arcs connecting their endpoints, and edges intersect at most once either in a proper crossing or in a shared endpoint. The edges and vertices of a drawing partition the plane into regions, which are called the \emph{cells} of the drawing. A \emph{triangle} in a simple drawing $D$ is a subdrawing of $D$ which is a drawing of~$K_3$. By the definition of simple drawings, any triangle is crossing free and thus splits the plane (or the sphere) in two connected regions. We call those regions the \emph{sides} of the triangle. If one side of a triangle does not contain any vertices of $D$, that side is called an \emph{empty side} of the triangle, and the triangle is called \emph{empty triangle}.
Note that empty (sides of) triangles might be intersected by edges. 
We observe that simple drawings of $K_3$ consist of exactly one triangle, which has two empty sides. 
Triangles in simple drawings of graphs with $n \geq4$ vertices have at most one empty side. 

In this work, we study the number of empty triangles in simple drawings of~$K_n$. 
Note that simple drawings 
are a topological generalization of straight-line drawings. 
The number $h_3(n)$ of empty triangles that every straight-line drawing of $K_n$ contains has been subject of intensive research.
It is easy to see that $h_3(n)=\Omega(n^2)$. The currently best known bounds are $n^2 - \frac{32}{7}n + \frac{22}{7} \leq h_3(n) \leq 1.6196n^2 + o(n^2)$~\cite{straight_triangle_lower,straight_triangle_upper}. 

For simple drawings of complete graphs, the situation changes drastically. Harborth~\cite{harborth} showed in 1989 that there are simple drawings of 
 $K_n$ that contain only $2n-4$ empty triangles; see \figurename~\ref{fig:gtwisted_k6}b. 
 This especially implies that most edges in these drawings 
 are not incident to any empty triangles. 
 On the other hand, Harborth observed that every vertex in these drawings  
 is incident to at least two empty triangles, a property 
 he 
 conjectured to be true in general. 
This conjecture has been proven 
in 2013~\cite{triangles_together,triangles_alone}. 
The currently best lower bound on the number of empty triangles in simple drawings of~$K_n$ is $n$~\cite{ahprsv-etgdc-15}. Further, it is conjectured that Harborth's upper bound should actually be the true lower bound.

\begin{conjct}[\cite{ahprsv-etgdc-15}]\label{conj:triangles}
	For any $n \geq 4$, every simple drawing of $K_n$ contains at~least \mbox{$2n-4$} empty triangles.
\end{conjct}

The drawings that Harborth used for his upper bound are now well known 
as \emph{twisted drawings}~\cite{bound_2003} 
and have received considerable attention~\cite{crossing_lemma,unavoidable_RS,biplanar_trees,used_twisted,notes_on_twisted,bound_2003,new_unavoidable}. 
A generalization of twisted drawings was introduced in~\cite{twisted_socg} as a special type of c-monotone drawings. 
A simple drawing~$D$ in the plane is \emph{{\cmonotone}} if there is a point~$O$ such that any ray emanating from~$O$ intersects any edge of~$D$ at most once. 
A c-monotone drawing $D$ is {\gtwisted} if there exists a ray~$r$ emanating from~$O$ that intersects every edge of~$D$.

As twisted drawings and the upper bound obtained by them are crucial in the study of empty triangles, it is natural to ask about the number of triangles in their generalization. 
The initial goal of this work was to prove Conjecture~\ref{conj:triangles} 
for generalized twisted drawings. 
As the number of such drawings is exponential (in the number of vertices), one might expect that they contain different numbers of empty triangles. 
However, we show that surprisingly, the conjectured bound is tight for all of them. 

\begin{theorem}\label{thm:number_empty}
	For any $n \geq 4$, every {\gtwisted} drawing of~$K_n$ 
	contains exactly $2n-4$ empty triangles.
\end{theorem}

\noindent {\bf Outline.\,} In Section~\ref{sec:pre}, we introduce some properties of generalized twisted drawings and empty triangles in general simple drawings. Then, in Section~\ref{sec:lemmata}, we show several results about empty triangles in generalized twisted drawings, which we finally put together to obtain a proof of Theorem~\ref{thm:number_empty}.

\section{Preliminaries}\label{sec:pre}

Two simple drawings are weakly isomorphic if the same pairs of edges cross. It is well-known that the weak isomorphism class of a drawing of $K_n$ completely determines which triangles are empty.
To prioritize readability, several of our figures show drawings that are weakly isomorphic to generalized twisted \mbox{(sub-)drawings} rather than a generalized twisted drawing.

\begin{figure}[tb]
	 	\centering
	 	\includegraphics[scale=0.8,page=9]{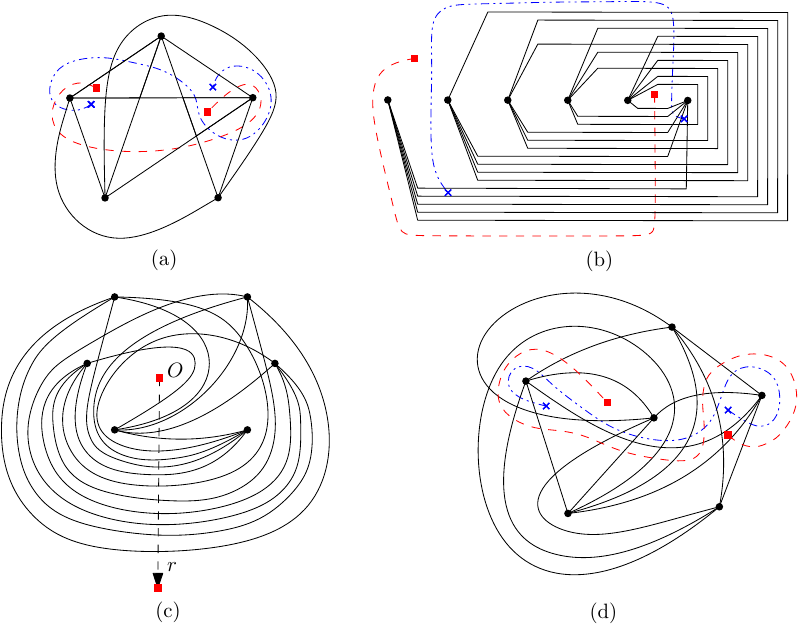}
		\caption{All (up to weak isomorphism) {\gtwisted} drawings of $K_5$ and~$K_6$. $O$ and $Z$ have to lie in cells marked with red squares or in cells with blue crosses. 
		Curves $OZ$ are drawn dashed or dash-dotted (and as ray in c).}
	 	\label{fig:gtwisted_k6}
\end{figure}

For a generalized twisted drawing~$D$ of $K_n$, we put a point $Z$ into the unbounded cell of $D$, 
on the ray~$r$ that crosses everything. 
Similarly, for
every drawing that is weakly isomorphic to a generalized twisted drawing,
there exists a simple curve $OZ$ corresponding to the part of the ray $r$ from $O$ to~$Z$; see~\cite{twisted_socg}.
Note that, given a simple drawing $D$ of $K_n$, there might be several cell pairs where $O$ and $Z$ could be placed such that $D$ is weakly isomorphic to a {\gtwisted} drawing with the 
corresponding cells for $O$ and $Z$. 
For instance, \figurename~\ref{fig:gtwisted_k6} shows all~\cite{twisted_socg} generalized twisted drawings of $K_5$ and $K_6$ up to weak isomorphism, together with all possible cell pairs for $O$ and $Z$ and some curve $OZ$ for each pair.
With this addition of $O$ and~$Z$, we will use the following properties of {\gtwisted} drawings, which have been shown in~\cite{twisted_socg}.

\begin{figure}[htb]
	 	\centering
	 	\includegraphics[scale=0.8,page=11]{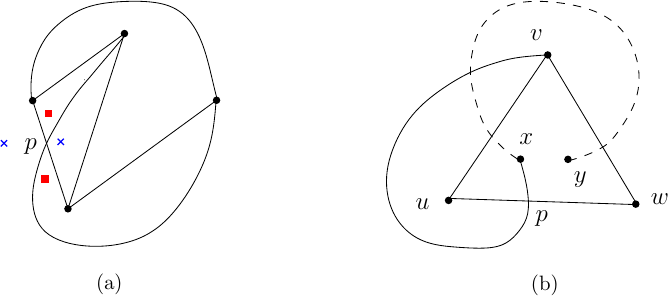}
		\caption{(a) $O$ and $Z$ have to lie in cells marked with red squares or in cells with blue crosses. (b) The edge $(x,y)$ cannot cross $(v,u)$ and $(v,w)$ simultaneously.}
	 	\label{fig:fourandfive}
\end{figure}

\begin{lemma}[\cite{twisted_socg}]\label{gtwisted_known}
	Let $D$ be a simple drawing in the plane that is weakly isomorphic to a {\gtwisted} drawing of $K_n$. Then the following holds:
	\begin{enumerate}
		\item\label{gtwisted_antipodal} For each triangle of $D$, the cell containing $O$ and the cell containing $Z$ lie on different sides. In particular, this implies that $D$ does not contain three interior-disjoint triangles.
		\item\label{has_to_be_vi} The cells containing $O$ and $Z$ 
		each have at least one vertex on their boundary.
		\item\label{gtwisted_maxcross} Every subdrawing of $D$ induced by four vertices contains a crossing. If $p$ is such a crossing, then $O$ and $Z$ lie in two different cells that are incident to~$p$ 
		and opposite to each other (see \figurename~\ref{fig:fourandfive}a for an illustration).
	\end{enumerate}
\end{lemma}

We will also use the following technical lemma for simple drawings.

\begin{lemma}\label{techical}
	Let $D$ be a simple drawing of $K_n$. Let $\Delta$ be a triangle of $D$ with vertices $u,v,w$. Let $x,y$ be two vertices on the same side of $\Delta$. If the edge $(x,v)$ crosses $(u,w)$, then the edge $(x,y)$ can cross at most one of $(v,u)$ and $(v,w)$.
\end{lemma}
\begin{proof}
Assume that $(x,y)$ crosses both $(v,u)$ and $(v,w)$. Since $x$ and $y$ are on the same side of $\Delta$, the edge $(x,y)$ must cross the boundary of $\Delta$ an even number of times. Thus, if $(x,y)$ crosses  $(v,u)$ and $(v,w)$, it cannot cross $(u,w)$. Let $p$ be the crossing point between $(x,v)$ and $(u,w)$.  See \figurename~\ref{fig:fourandfive}b for an illustration. Suppose that $(x,y)$ crosses first $(v,u)$ and then $(v,w)$. After crossing $(v,u)$, the edge $(x,y)$ and the vertex $y$ are in different regions defined by the closed curve~$C$ consisting of $(v,u)$, the part of the edge $(u,w)$ from $u$ to $p$ and the part of the edge $(x,v)$ from $p$ to $v$. Then, after crossing $(v,u)$, the edge $(x,y)$ must cross $C$ to reach $y$, and this is not possible without violating the simplicity of the drawing. Therefore, $(x,y)$ cannot cross $(v,u)$ and $(v,w)$ simultaneously. 
An analogous analysis can be done if $(x,y)$ crosses first $(v,w)$ and then $(v,u)$.
\end{proof}

In addition to the properties of generalized twisted drawings, we will use the concept of star triangles as introduced in~\cite{ahprsv-etgdc-15}. A triangle $\Delta$ with vertices $x,y,z$ is a \emph{star triangle} at $x$ if $yz$ is not crossed by any edges incident to~$x$. We will use the following properties of star triangles in simple drawings of~$K_n$.

\begin{lemma}\label{lem:gen_star}
	Let $D$ be a simple drawing of $K_n$ in the plane and $x$ be a vertex of~$D$. Then the following holds:
	\begin{enumerate}
		\item\label{gen_two} There are at least two empty star triangles at~$x$.
		\item\label{gen_consec} A star triangle $xyz$ at a vertex~$x$ is an empty triangle if and only if the vertices $y$ and $z$ are consecutive in the rotation around~$x$. 
		\item\label{gen_disjoint} For any two different empty star triangles at $x$, $xy'z'$ and $xyz$, the empty sides of $xy'z'$ and $xyz$ are disjoint.
	\end{enumerate}
\end{lemma}

\begin{proof}

Properties \ref{gen_two} and \ref{gen_consec} have been shown in~\cite{ahprsv-etgdc-15}. 	To prove Property~\ref{gen_disjoint}, consider the boundary edges of the triangles $xyz$ and $xy'z'$. 
Of these edges, the only pair that could cross is $(y,z)$ and $(y',z')$. However, $y'$ and $z'$ lie on the same side of the triangle $xyz$, so $(y',z')$ has to cross the boundary of $xyz$ an even number of times, which is not possible if exactly $(y,z)$ and $(y',z')$ cross. Thus, no edges on the boundary of the star triangles cross, and therefore their empty sides are disjoint.
\end{proof}

\section{Proof of Theorem~\ref{thm:number_empty}}\label{sec:lemmata}
In the following, we derive several lemmata about empty triangles in {\gtwisted} drawings. 
These lemmata put together will give the proof of Theorem~\ref{thm:number_empty}. 

\begin{lemma}\label{prop:exactly_2_star_triangles}
		Let $D$ be a {\gtwisted} drawing of~$K_n$ in the plane with~$n\geq 4$ and $x$ be a vertex of $D$. Then $x$ is incident to exactly two empty star triangles, one 
		has $O$ on the empty side and the other has $Z$ on the empty side. 
		Further, these star triangles have disjoint empty sides. 
\end{lemma}

\begin{proof}
	By Lemma~\ref{lem:gen_star}(\ref{gen_two} and \ref{gen_disjoint}), for every vertex~$x$ there are at least two empty star triangles at~$x$ and the empty sides of these triangles are disjoint. By Lemma~\ref{gtwisted_known}(\ref{gtwisted_antipodal}), any triangle of a {\gtwisted} drawing has $O$ on one side and $Z$ on the other side, and $D$ cannot contain three interior-disjoint triangles. Thus, for any vertex $x$ in a {\gtwisted} drawing it holds that: (i) one empty star triangle at $x$ 
	has $O$ on the empty side, (ii) another empty star triangle at $x$
	has $Z$ on the empty side, and (iii) there cannot be a third empty star triangle at~$x$.
\end{proof}

\begin{lemma}\label{prop:v1_triangle}
	Let $D$ be a {\gtwisted} drawing of~$K_n$ with~$n\geq 4$. Let $C_O$ be the cell of $D$ containing~$O$ and let $v$ be a vertex on the boundary of~$C_O$. 
	Let 
	$\Delta$
	be an empty triangle in $D$ that has $O$ on the empty side. 
		Then the following holds:
	\begin{enumerate}
		\item \label{emptyv} The vertex $v$ is a vertex of $\Delta$, that is, $\Delta=xyv$ for some $x,y$. 
	\item \label{empty1or2}
		The triangle $\Delta=xyv$ is an empty star triangle at $x$ or $y$ or both.
	\item \label{emptyno3} 
		If $\Delta=xyv$ is a star triangle at both $x$ and $y$,  
		then all edges emanating from 
		$v$ cross $(x,y)$.
			Hence, $\Delta$ 
			is a star triangle for at most two of its vertices.  
	\end{enumerate}
\end{lemma}
\begin{proof}
	Since $\Delta$ has $O$ on its empty side and since $C_O$ is a cell of $D$, $\Delta$ also has $C_O$ on the empty side.
	Therefore, since $v$ belongs to $C_O$ and $\Delta$ is empty, necessarily $v$ must be one of the vertices of $\Delta$ and Property~\ref{emptyv} is fulfilled. 

To prove Property~\ref{empty1or2}, assume to the contrary that $\Delta = xyv$ is not a star triangle at $x$ or $y$. Then at least one 
edge $(x,x')$ must cross $(v,y)$ at a point $q'$ and at least an edge $(y,y')$ must cross $(v,x)$ at a point $q$ (see \figurename~\ref{fig:starvertices}a for 
an illustration). Note that since $\Delta$ is empty and any edge incident to $x$ or $y$ can cross $\Delta$ at most once, $(x,x')$ and $(y,y')$ must emanate from $x$ and $y$, respectively, at the empty side of $\Delta$, and cross at a point $p$ on that side of $\Delta$. Without loss of generality, we may assume that $O$ is very close to $v$.
Consider the subdrawing $D'$ induced by $x,x',y$ and $y'$. Observe that since $(x,y')$ cannot cross $(x,v)$ or $(y,y')$,  the cell of $D'$ defined by $x,p$ and $y'$ cannot contain $O$, regardless of the shape of $(x,y')$. 
Thus, by Lemma~\ref{gtwisted_known}(\ref{gtwisted_maxcross}) applied to $D'$, $O$ is in the region defined by $x', p$ and $y'$, and $Z$ is in the region defined by $x, p$ and $y$, contradicting that $O$ and $Z$ lie on different sides of $\Delta$.

\begin{figure}[htb]
	 	\centering
	 	\includegraphics[scale=0.8,page=3]{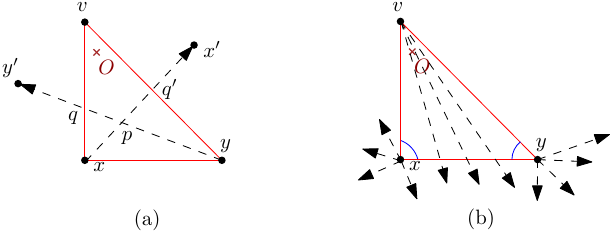}
		\caption{(a) Illustrating the proof of Lemma~\ref{prop:v1_triangle}(\ref{empty1or2}).
		(b) Any empty triangle of $D$ that has $O$ on the empty side cannot be a star triangle at three vertices.}
	 	\label{fig:starvertices}
\end{figure}

To prove Property~\ref{emptyno3}, take a vertex $w$ that 
is not a vertex of~$\Delta$. By Lemma~\ref{gtwisted_known}(\ref{gtwisted_maxcross}), the subdrawing induced by $v, x, y$ and $w$ has a crossing. As $\Delta$ is a star triangle at $x$ and $y$, any edge incident to $x$ or $y$ emanates from $x$ or $y$ on the non-empty side of $\Delta$, so neither $(x,w)$ nor $(y,w)$ can cross $\Delta$. Then $(v,w)$ must cross $(x,y)$, emanating from $v$ on the empty side of $\Delta$. Therefore, Property~\ref{emptyno3} follows.
\end{proof}

Note that by Lemma~\ref{gtwisted_known}(\ref{has_to_be_vi}), the cell containing $O$ always has a vertex on its boundary. 
Hence, by Lemma~\ref{prop:v1_triangle}, any empty triangle with $O$ on the empty side is a star triangle at one or two vertices.
The following lemma proves that there are exactly two such triangles that are star triangles at two vertices.

\begin{lemma}\label{prop:extra_type_2_triangle}
	Let $D$ be a {\gtwisted} drawing of~$K_n$ with~$n\geq 4$. Then $D$ contains exactly two empty triangles with $O$ on the empty side that are star triangles at two vertices. 
\end{lemma}
\begin{proof}
	Let $C_O$ be the cell containing $O$ and $v$ be a vertex on the boundary of~$C_O$, which exists by Lemma~\ref{gtwisted_known}(\ref{has_to_be_vi}). By Lemma~\ref{prop:exactly_2_star_triangles}, there is an empty star triangle $\Delta =vuw$ at $v$ that has $C_O$ on the empty side. By Lemma~\ref{prop:v1_triangle}, $\Delta$ is a star triangle at exactly one of $u$ or $w$, say $w$. Thus, $\Delta$ is an empty star triangle at two vertices with $O$ on the empty side.

On the other hand, 
	by Lemma~\ref{prop:v1_triangle}(\ref{emptyno3}), all edges emanating from $u$ cross $(v,w)$.
	Among all edges crossing $(v,w)$, we choose the edge $(a,b)$ that crosses $(v,w)$ closest to $v$. 
	Let $p$ be the crossing point between $(a,b)$ and $(v,w)$, and consider the triangle~$vab$. 
	We will show that the triangle $vab$ is a star triangle at $a$ and $b$ and that the side $F$ of it in which $O$ lies is empty.  

	If $a\neq u$, then $(a,b)$ crosses $(v,u)$ at a point $q$; see \figurename~\ref{fig:case_only_v1}.  
Note that $F$ is partitioned into three triangular shapes $vaq$, $vqp$, and $vpb$, where $O$ lies in $vqp$. 
Assume, for a contradiction, that a vertex $x$ lies in $vbp$. Since $(a,x)$ can cross neither $(a,b)$ nor $(v,p)$, it has to cross $(v,b)$. As no simple drawing of the $K_4$ can contain more than one crossing, the edges $xv$ and $xb$ have to stay completely in $F$. Since $(v,p)$ is crossing-free, the edges $(x,v)$ and $(x,b)$ must be in $vbp$, one side of the triangle $vxb$ is contained in $vpb$. As $O$ is not on the side of $vxb$ contained in $F$, then $Z$ has to be on that side. This implies that both $O$ and $Z$ are in $F$, a contradiction. Therefore, $vbp$ is empty. Using a similar argument, one can prove that $vqa$ is also empty, so $F$ is empty. Besides, any edge incident to $a$ or $b$ must emanate outside $F$ because $(v,p)$ is crossing free. As a consequence, $F$ is the empty side of the triangle $vab$, which is a star triangle at $a$ and $b$. 

	The reasoning for $a=u$ is similar (with two triangular shapes $vup$ and $vpb$).
	
	 \begin{figure}[htb]
	 	\centering
	 	\includegraphics[scale=0.8,page=5]{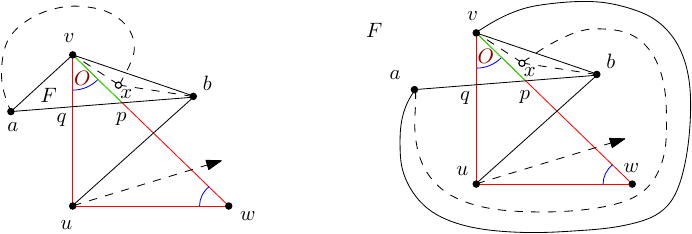}
	 	\caption{
	 	Illustrating the proof of Lemma~\ref{prop:extra_type_2_triangle}.}
	 	\label{fig:case_only_v1}
	 \end{figure}

	 What remains to show is that there is no third empty triangle with $O$ on the empty side that is a star triangle at two vertices. Assume for a contradiction that such a triangle $x'y'z'$ exists. By Lemma~\ref{prop:v1_triangle}(\ref{emptyv}), one of $x', y'$ and $z'$ must be $v$, say $z'=v$. As $v$, $a$, and $b$ are incident to at most one empty star triangle with $O$ on the empty side, by Lemma~\ref{prop:exactly_2_star_triangles}, $x'$ and $y'$ are different from $a$ and $b$, and $vx'y'$ is a star triangle at $x'$ and $y'$. Consider the triangle $vab$. Since $x'$ and $y'$ are on the same side of $vab$ and $(v,x')$ crosses $(a,b)$ by Lemma~\ref{prop:v1_triangle}(\ref{emptyno3}), the edge $(x',y')$ cannot cross $(v,a)$ and $(v,b)$ by Lemma~\ref{techical}. But 
	 all edges emanating from $v$ must cross $(x',y')$ by Lemma~\ref{prop:v1_triangle}(\ref{emptyno3}), a contradiction. Hence $x'y'z'$ cannot exist.
\end{proof}

We note that the lemmata above and their proofs hold for every choice of~$C_O$ (if there are many) and any vertex $v$ on the boundary $C_O$. However, whether a triangle is empty and at how many vertices it is a star triangle does not change between weakly isomorphic drawings. As a consequence, the empty star triangles obtained in the previous lemmata and proofs must be the same, regardless of the choice of $C_O$ and the vertex $v$ on the boundary of~$C_O$. We also note that for empty triangles having $Z$ 
on the empty side, the reasoning in Lemmas~\ref{prop:v1_triangle} and~\ref{prop:extra_type_2_triangle} works to prove that these triangles are star triangles at one or two vertices, and that exactly two of them are star triangles at two vertices. By Lemma~\ref{prop:extra_type_2_triangle} and these observations, we get the following lemma.

\begin{lemma}\label{cor:extra_type_2_triangle}
Let $D$ be a {\gtwisted} drawing of~$K_n$ with~$n\geq 4$. Then $D$ contains exactly four empty triangles that are star triangles at two vertices.	
\end{lemma}	

Now, we can prove our main theorem.

\begin{proof}[of Theorem~\ref{thm:number_empty}]
	When summing up the number of empty star triangles over all vertices, we obtain $2n$ empty star triangles by Lemma~\ref{prop:exactly_2_star_triangles} ($n$ triangles with $O$ on the empty side and $n$ with $Z$ on the empty side). By Lemma~\ref{prop:v1_triangle}, all empty triangles have been counted this way, but the triangles that are empty star triangles at two vertices have been counted twice. By Lemma~\ref{cor:extra_type_2_triangle}, there are exactly four triangles that are empty star triangles at two vertices. Thus, there are exactly four triangles that have been counted exactly twice and the precise number of empty triangles in $D$ is $2n-4$.
\end{proof}

\bibliography{gd_empty_triangles_in_gtwisted_drawings}

\begin{thebibliography}{10}

\bibitem{crossing_lemma}
Bernardo {\'{A}}brego, Silvia Fern{\'{a}}ndez-Merchant, Ana~Paulina Figueroa,
  Juan~Jos{\'{e}} Montellano-Ballesteros, and Eduardo Rivera-Campo.
\newblock Crossings in twisted graphs.
\newblock In {\em Collection of Abstracts for the 22nd Japan Conference on
  Discrete and Computational Geometry, Graphs, and Games (JCDCG3)}, pages
  21--22, 2019.
\newblock URL:
  \url{http://www.csun.edu/~sf70713/publications/2019_JCDCG3_Twisted_Crossings.pdf}.

\bibitem{straight_triangle_lower}
Oswin Aichholzer, Ruy Fabila-Monroy, Thomas Hackl, Clemens Huemer, Alexander
  Pilz, and Birgit Vogtenhuber.
\newblock Lower bounds for the number of small convex {$k$}-holes.
\newblock {\em Computational Geometry. Theory and Applications},
  47(5):605--613, 2014.
\newblock \href {https://doi.org/10.1016/j.comgeo.2013.12.002}
  {\path{doi:10.1016/j.comgeo.2013.12.002}}.

\bibitem{twisted_socg}
Oswin Aichholzer, Alfredo Garc\'{\i}a, Javier Tejel, Birgit Vogtenhuber, and
  Alexandra Weinberger.
\newblock Twisted ways to find plane structures in simple drawings of complete
  graphs.
\newblock In {\em Proceedings of the 38th International Symposium on
  Computational Geometry (SoCG 2022)}, pages 5:1--5:18, 2022.
\newblock \href {https://doi.org/10.4230/LIPIcs.SoCG.2022.5}
  {\path{doi:10.4230/LIPIcs.SoCG.2022.5}}.

\bibitem{ahprsv-etgdc-15}
Oswin Aichholzer, Thomas Hackl, Alexander Pilz, Pedro Ramos, Vera
  Sacrist\'{a}n, and Birgit Vogtenhuber.
\newblock Empty triangles in good drawings of the complete graph.
\newblock {\em Graphs and Combinatorics}, 31(2):335--345, 2015.
\newblock \href {https://doi.org/10.1007/s00373-015-1550-5}
  {\path{doi:10.1007/s00373-015-1550-5}}.

\bibitem{unavoidable_RS}
Alan Arroyo, R.~Bruce Richter, Gelasio Salazar, and Matthew Sullivan.
\newblock The unavoidable rotation systems, 2019.
\newblock URL: \url{https://arxiv.org/abs/1910.12834}.

\bibitem{straight_triangle_upper}
Imre B{\'a}r{\'a}ny and Pavel Valtr.
\newblock Planar point sets with a small number of empty convex polygons.
\newblock {\em Studia Scientiarum Mathematicarum Hungarica. A Quarterly of the
  Hungarian Academy of Sciences}, 41:243--266, 2004.
\newblock \href {https://doi.org/10.1556/SScMath.41.2004.2.4}
  {\path{doi:10.1556/SScMath.41.2004.2.4}}.

\bibitem{biplanar_trees}
Ana~Paulina Figueroa and Juli{\'{a}}n Fres{\'{a}}n-Figueroa.
\newblock The biplanar tree graph.
\newblock {\em Boletín de la Sociedad Matemática Mexicana}, 26:795--806,
  2020.
\newblock \href {https://doi.org/10.1007/s40590-020-00287-y}
  {\path{doi:10.1007/s40590-020-00287-y}}.

\bibitem{triangles_together}
Radoslav Fulek and Andres~J. Ruiz-Vargas.
\newblock Topological graphs: empty triangles and disjoint matchings.
\newblock In {\em Proceedings of the 29th Annual Symposium on Computational
  Geometry (SoCG'13)}, pages 259--266, New York, 2013. ACM.
\newblock \href {https://doi.org/10.1145/2462356.2462394}
  {\path{doi:10.1145/2462356.2462394}}.

\bibitem{harborth}
Heiko Harborth.
\newblock Empty triangles in drawings of the complete graph.
\newblock {\em Discrete Mathematics}, 191:109--111, 1998.

\bibitem{used_twisted}
Jan Kyn{\v{c}}l and Pavel Valtr.
\newblock On edges crossing few other edges in simple topological complete
  graphs.
\newblock {\em Discrete Mathematics}, 309(7):1917--1923, 2009.
\newblock 13th International Symposium on Graph Drawing, 2005.
\newblock \href {https://doi.org/10.1016/j.disc.2008.03.005}
  {\path{doi:10.1016/j.disc.2008.03.005}}.

\bibitem{notes_on_twisted}
Elsa Oma{\~{n}}a-Pulido and Eduardo Rivera-Campo.
\newblock Notes on the twisted graph.
\newblock In {\em Computational Geometry: XIV Spanish Meeting on Computational
  Geometry, EGC 2011, Dedicated to Ferran Hurtado on the Occasion of His 60th
  Birthday, Alcal{\'a} de Henares, Spain, June 27-30, 2011, Revised Selected
  Papers}, volume 7579 of {\em Lecture Notes in Computer Science (LNCS)}, pages
  119--125. 2012.

\bibitem{bound_2003}
J\'{a}nos Pach, J\'{o}zsef Solymosi, and G{\'e}zak T\'{o}th.
\newblock Unavoidable configurations in complete topological graphs.
\newblock {\em Discrete Comput Geometry}, 30:311--320, 2003.
\newblock \href {https://doi.org/10.1007/s00454-003-0012-9}
  {\path{doi:10.1007/s00454-003-0012-9}}.

\bibitem{triangles_alone}
Andres~J. Ruiz-Vargas.
\newblock Empty triangles in complete topological graphs.
\newblock In {\em Discrete Computational Geometry}, volume~53, pages 703--712,
  2015.
\newblock \href {https://doi.org/10.1007/s00454-015-9671-4}
  {\path{doi:10.1007/s00454-015-9671-4}}.

\bibitem{new_unavoidable}
Andrew Suk and Ji~Zeng.
\newblock Unavoidable patterns in complete simple topological graphs, 2022.
\newblock URL: \url{https://arxiv.org/abs/2204.04293}.

\end{thebibliography}
\end{document}